# A Unified Description for Polarization-Transfer Mechanisms in Magnetic Resonance in Static Solids: Cross Polarization and DNP

Zhenfeng Pang[a,d], Sheetal Jain[,b], Chen Yang[c], Xueqian Kong[a], and Kong Ooi Tan[d]*

[a]Department of Chemistry, Zhejiang University, 310027 Hangzhou, China
[b]Solid State and Structural Chemistry Unit, Indian Institute of Science, Bangalore 560012, India
[c]Amazon Robotics, 300 Riverpark Drive, North Reading, Massachusetts 01864, United States
[d]Laboratoire des Biomolécules, LBM, Département de Chimie, École Normale Supérieure, PSL University, Sorbonne Université, CNRS, 75005 Paris, France

*Corresponding Authors:

Kong Ooi Tan
Laboratoire des Biomolécules, LBM, Département de Chimie,
École Normale Supérieure, PSL University, Sorbonne Université, CNRS,
75005 Paris, France
ORCID:0000-0002-3094-3398
E-Mail: kong-ooi.tan@ens.psl.eu

Abstract

Polarization transfers are crucial building blocks in magnetic resonance experiments, i.e., they can be used to polarize insensitive nuclei and correlate nuclear spins in multidimensional NMR spectroscopy. The polarization can be transferred either across different nuclear spin species or from electron spins to the relatively low-polarized nuclear spins. The former route occurring in solid-state NMR (ssNMR) can be performed via cross polarization (CP), while the latter route is known as dynamic nuclear polarization (DNP). Despite having different operating conditions, we opinionate that both mechanisms are theoretically similar processes in ideal conditions, i.e., the electron is merely another spin-1/2 particle with a much higher gyromagnetic ratio. Here, we show that the CP and DNP processes can be described using a unified theory based on average Hamiltonian theory (AHT) combined with fictitious operators. The intuitive and unified approach has allowed new insights on the cross effect (CE) DNP mechanism, leading to better design of DNP polarizing agents and extending the applications beyond just hyperpolarization. We explore the possibility of exploiting theoretically predicted DNP transients for electron-nucleus distance measurements—like routine dipolar-recoupling experiments in solid-state NMR.

# 1 Introduction

Polarization-transfer experiments play crucial roles in magnetic resonance spectroscopy. Not only do they enhance the sensitivity of the insensitive nuclei, but also allow distance measurement and, hence, structure determination in the system of interest. Two examples of such experiments are cross polarization (CP),[1] which facilitates polarization transfer via a spin-locking technique in solid-state NMR, and dynamic nuclear polarization (DNP), which enables the transfer from unpaired electrons to nuclei mediated by strategic microwave (μw) irradiation.[2,3] The CP mechanism was first theoretically explained using a spin-thermodynamic approach,[1] which predicts an exponential time dependence of nuclear polarization during the buildup. However, the semi-classical treatment was then shown to be inconsistent with the observation of the transient oscillations in a ferrocene single crystal—a phenomenon that was accurately described using the product-operator formalism that adopts a quantum-mechanical approach.[4,5] The transient oscillation was exploited in many NMR experiments to measure the distance between nuclear spins accurately.[6–8]

Although the analytical theory has long been developed for explaining DNP mechanisms, in particular, solid effect (SE) and cross effect (CE),[9–11] they are primarily adapted for cases when energy-level transitions are saturated with low-power continuous-wave (CW) microwaves. In such situations, perturbation theory is applied to theoretically describe the nuclear polarizations during DNP. [12–15] Nevertheless, we expect such a treatment to be less appropriate when strong μw powers are applied. For instance, it might fail to describe transient oscillations, a common phenomenon in many polarization-transfer experiments, due to poor convergence to exact numerical solutions.[4] Therefore, there is a need to review the DNP theory that applies to high-power μw conditions. Microwave instrumentation has been rapidly developing since the commercialization of magic-angle spinning (MAS) DNP and dissolution DNP (D-DNP) spectrometers.[16–18] With this rapid progress, high-field pulsed DNP spectroscopy [19] may become available in the near future.

In this work, we show that CP, NOVEL (Nuclear spin Orientation Via Electron spin Locking), SE, and CE (Fig. 1) can be explained with an analytical theory based on average Hamiltonian theory (AHT) combined with fictitious operators in subspaces.[20–23] Thus, we opinionate that DNP and CP are fundamentally similar processes in terms of spin physics in ideal situations, i.e., the relaxation rates are negligible so that both processes can be described using a unified theoretical framework. The exact analytical results obtained from the unified theory will shed new light on CE mechanism, which could help design better DNP polarizing agents, and further extend the DNP applications beyond just hyperpolarization, i.e., measuring electron-nucleus correlations, distances, and relative orientations for structure determination in biological molecules or materials.

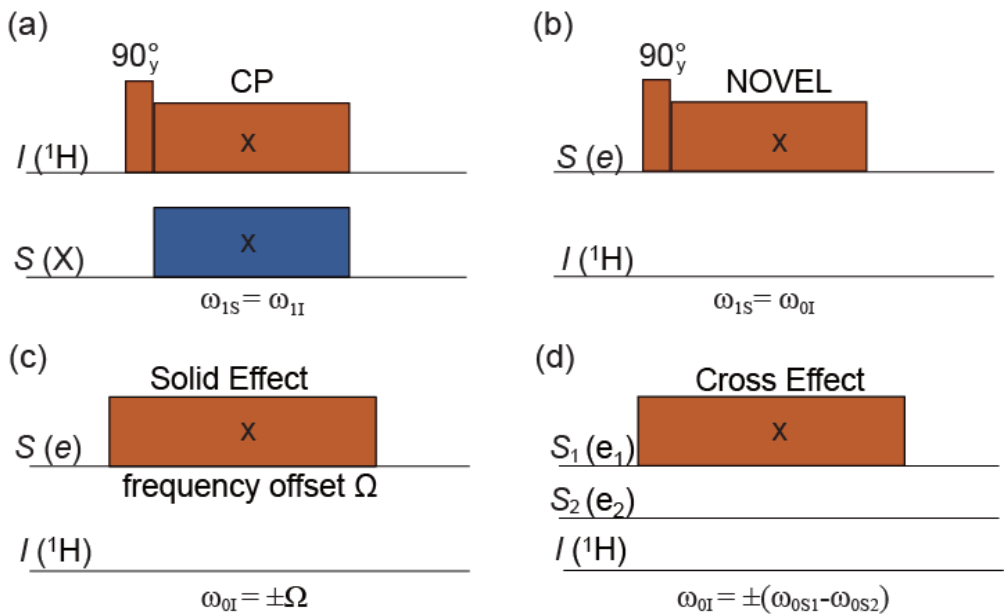

Fig. 1 Schematic diagrams of CP, NOVEL, SE, and CE pulse sequences and matching conditions.

## 2 Theory

### 2.1 Cross polarization (CP)

We begin by first writing down the Hamiltonian of a two-spin $IS$ system in the double rf rotating frame:

$$\widehat{\mathcal{H}}_{CP} = 2d_{IS}\hat{S}_z\hat{I}_z + \omega_{1S}\hat{S}_x + \omega_{1I}\hat{I}_x, \tag{1}$$

where $\omega_{1S}$ and $\omega_{1I}$ are the nutation frequencies of the $S$ and $I$ spin along the x-axes, respectively; $d_{IS}$ denotes the dipolar coupling between the two spins. The frame is then rotated with a propagator $\widehat{U}_t = \exp(-i(\pi \hat{S}_y + \pi \hat{I}_y)/2)$ so that the z-axis[1] is now defined along with the rf fields:

$$\widehat{\mathcal{H}}_t = \widehat{U}_t^{-1} \widehat{\mathcal{H}}_{CP} \widehat{U}_t \qquad (2)$$
$$= 2d_{IS}\hat{S}_x\hat{I}_x + \omega_{1S}\hat{S}_z + \omega_{1I}\hat{I}_z,$$

followed by another interaction-frame transformation using $\widehat{U}_1(t) = \exp(-i(\omega_{1S}\hat{S}_z + \omega_{1I}\hat{I}_z)t)$:

$$\widehat{\widetilde{\mathcal{H}}}(t) = \widehat{U}_1^{-1}(t)\widehat{\mathcal{H}}_t\widehat{U}_1(t) - (\omega_{1S}\hat{S}_z + \omega_{1I}\hat{I}_z)$$
$$= d_{IS}(\hat{S}_x\hat{I}_x + \hat{S}_y\hat{I}_y)\cos(\omega_{1S} - \omega_{1I})t + d_{IS}(\hat{S}_x\hat{I}_y - \hat{S}_y\hat{I}_x)\sin(\omega_{1S} - \omega_{1I})t \qquad (3)$$
$$+ d_{IS}(\hat{S}_x\hat{I}_x - \hat{S}_y\hat{I}_y)\cos(\omega_{1S} + \omega_{1I})t - d_{IS}(\hat{S}_x\hat{I}_y + \hat{S}_y\hat{I}_x)\sin(\omega_{1S} + \omega_{1I})t.$$

Then, $\widehat{\widetilde{\mathcal{H}}}(t)$ becomes time-independent if we apply $\omega_{1S} = \omega_{1I}$ (Hartmann-Hahn condition) and AHT in the second step:

$$\widehat{\widetilde{\mathcal{H}}}(t) = d_{IS}[(\hat{S}_x\hat{I}_x + \hat{S}_y\hat{I}_y) + (\hat{S}_x\hat{I}_x - \hat{S}_y\hat{I}_y)\cos 2\omega_{1I}t - (\hat{S}_x\hat{I}_y + \hat{S}_y\hat{I}_x)\sin 2\omega_{1I}t], \qquad (4)$$

$$\widehat{\widetilde{\mathcal{H}}} = \frac{\omega_{1I}}{\pi}\int_0^{\pi/\omega_{1I}} \widehat{\widetilde{\mathcal{H}}}(t)\,dt = d_{IS}(\hat{S}_x\hat{I}_x + \hat{S}_y\hat{I}_y). \qquad (5)$$

Note that the AHT treatment is valid if the chosen cycle time ($\tau_c = \pi/\omega_{1I}$) is short compared to $\sim 2\pi/d_{IS}$. After that, the evolution of the initial density operator, $\hat{\rho}(0) = \hat{I}_z$, in the spin-locked frame under $\widehat{\widetilde{\mathcal{H}}}$ (Eq. 5) can be computed using the Liouville von Neumann (LvN) equation:

$$\hat{\rho}(t) = \widehat{U}_{CP}\hat{\rho}(0)\widehat{U}_{CP}^{-1} = \hat{I}_z\cos^2(\omega_{CP}t) + \hat{S}_z\sin^2(\omega_{CP}t) + (\hat{S}_y\hat{I}_x - \hat{S}_x\hat{I}_y)\sin(2\omega_{CP}t), \qquad (6)$$

where $\omega_{CP} = d_{IS}/2$ and $\widehat{U}_{CP} = \exp(-i\widehat{\widetilde{\mathcal{H}}}t)$. It is evident that the first two terms in Eq. 6 show that the polarization has been transferred from $\hat{I}_z$ to $\hat{S}_z$; the transfer was mediated by $\hat{S}_x^\Delta = \hat{S}_x\hat{I}_x + \hat{S}_y\hat{I}_y$, a familiar fictitious spin-1/2 operator in the ZQ subspace.[5] By realizing other fictitious operators—including those in the ZQ and double-quantum (DQ) subspaces—and their commutator relations,[2] one can describe the transfer in a more compact form:

$$\hat{I}_z = \hat{S}_z^\Sigma - \hat{S}_z^\Delta \xrightarrow{2\omega_{CP}\hat{S}_x^\Delta} \hat{S}_z^\Sigma - \hat{S}_z^\Delta \cos(2\omega_{CP}t) + \hat{S}_y^\Sigma \sin(2\omega_{CP}t)$$

$$\to \hat{S}_z^\Sigma + \hat{S}_z^\Delta = \hat{S}_z \text{ if } t = \pi/(2\omega_{CP}). \tag{7}$$

---

[1] Suppose we transfer the polarization from $I$ to $S$, the initial density matrix during spin lock is $\hat{\rho}(0) = \hat{I}_z$ in the double rotating frame. The process of polarization transfer can be easier analyzed in this frame.

Thus, we have exemplified here that using the fictitious operators in AHT offers a simple yet insightful approach to understanding CP. Note that the realization and identification of these fictitious operators in the subspaces are important elements to describe DNP processes (vide infra).

## 2.2  NOVEL

NOVEL is often referred to as the CP-equivalent sequence in DNP due to their similar matching conditions,[20,24] i.e., the electron Rabi field, $\omega_{1S}$, during spin-lock is set to match the nuclear Larmor frequency, $\omega_{0I}$ in NOVEL (Fig. 1b). The Hamiltonian of a two-spin electron-nucleus system during the spin-lock in the electron-rotating-frame is given by:

$$\hat{\mathcal{H}}_{NOVEL} = -\omega_{0I}\hat{I}_z + A_{zz}\hat{S}_z\hat{I}_z + A_{zx}\hat{S}_z\hat{I}_x + A_{zy}\hat{S}_z\hat{I}_y + \omega_{1S}\hat{S}_x, \tag{8}$$

where $A_{zz}$ and $A_{zx(y)}$ are the secular and pseudo-secular components of the hyperfine interaction. Then, a tilted-frame transformation using $\hat{U}_s = \exp(-i\varphi\hat{I}_z)$ and $\varphi = \tan^{-1}(A_{zy}/A_{zx})$ is performed along $\hat{I}_z$ to obtain:

$$\hat{\mathcal{H}}_s = \hat{U}_s^{-1}\hat{\mathcal{H}}_{NOVEL}\hat{U}_s = -\omega_{0I}\hat{I}_z + A_{zz}\hat{S}_z\hat{I}_z + B_{zx}\hat{S}_z\hat{I}_x + \omega_{1S}\hat{S}_x, \tag{9}$$

where $B_{zx} = \sqrt{A_{zx}^2 + A_{zy}^2}$. Following a similar treatment shown in CP (section 2.1), a propagator $\hat{U}_t = \exp(-i\pi\hat{S}_y/2)$ is applied to set the electron z-axis along the µw spin-lock field:

---

[2] Other ZQ operators are $\hat{S}_z^\Delta = (\hat{S}_z - \hat{I}_z)/2$ and $\hat{S}_y^\Delta = \hat{S}_y\hat{I}_x - \hat{S}_x\hat{I}_y$, and the relevant DQ operator here is $\hat{S}_z^\Sigma = (\hat{S}_z + \hat{I}_z)/2$. These operators obey $[\hat{S}_x^\Delta, \hat{S}_y^\Delta] = i\hat{S}_z^\Delta$ and $[\hat{S}_{x,y,z}^\Delta, \hat{S}_{x,y,z}^\Sigma] = 0$.

$$\widehat{\mathcal{H}}_t = \widehat{U}_t^{-1} \widehat{\mathcal{H}}_s \widehat{U}_t = \omega_{1S} \hat{S}_z - \omega_{0I} \hat{I}_z - A_{zz} \hat{S}_x \hat{I}_z - B_{zx} \hat{S}_x \hat{I}_x, \tag{10}$$

followed by another interaction-frame transformation using $\widehat{U}_1 = \exp(-i(\omega_{1S}\hat{S}_z - \omega_{0I}\hat{I}_z)t)$:

$$\begin{aligned}\widehat{\widetilde{\mathcal{H}}}(t) &= \widehat{U}_1^{-1}(t)\widehat{\mathcal{H}}_t \widehat{U}_1(t) - (\omega_{1S}\hat{S}_z - \omega_{0I}\hat{I}_z)\\ &= -A_{zz}(\cos\omega_{1S}t\,\hat{S}_x - \sin\omega_{1S}t\,\hat{S}_y)\hat{I}_z \\ &\quad -B_{zx}(\cos\omega_{1S}t\,\hat{S}_x - \sin\omega_{1S}t\,\hat{S}_y)(\cos\omega_{0I}t\,\hat{I}_x + \sin\omega_{0I}t\,\hat{I}_y).\end{aligned} \tag{11}$$

Similarly, $\widehat{\widetilde{\mathcal{H}}}(t)$ becomes time-independent if one sets $\omega_{1S} = \pm\omega_{0I}$ and apply AHT in the second step:

$$\begin{aligned}\widehat{\widetilde{\mathcal{H}}}(t) =\; & -\frac{B_{zx}}{2}\bigl(\hat{S}_x\hat{I}_x \mp \hat{S}_y\hat{I}_y\bigr) \\ & -\frac{B_{zx}}{2}\bigl[(\hat{S}_x\hat{I}_x \pm \hat{S}_y\hat{I}_y)\cos 2\omega_{0I}t + (\hat{S}_x\hat{I}_y \mp \hat{S}_y\hat{I}_x)\sin 2\omega_{0I}t\bigr] \\ & -A_{zz}(\cos\omega_{0I}\,\hat{S}_x\hat{I}_z \mp \sin\omega_{0I}t\,\hat{S}_y\hat{I}_z).\end{aligned} \tag{12}$$

$$\begin{aligned}\widehat{\overline{\mathcal{H}}} &= \frac{\omega_{0I}}{2\pi}\int_0^{2\pi/\omega_{0I}} \widehat{\widetilde{\mathcal{H}}}(t)\, dt \\ &= \begin{cases} -\frac{B_{zx}}{2}(\hat{S}_x\hat{I}_x + \hat{S}_y\hat{I}_y) & \text{if } \omega_{1S} = -\omega_{0I}\,(\text{ZQ}) \\ -\frac{B_{zx}}{2}(\hat{S}_x\hat{I}_x - \hat{S}_y\hat{I}_y) & \text{if } \omega_{1S} = \omega_{0I}\,(\text{DQ}) \end{cases}.\end{aligned} \tag{13}$$

Note that the AHT treatment is valid if $\omega_{0I} \gg A_{zz}, B_{zx}$, which is true for most DNP experiments at high fields. Also, one can obtain ZQ (DQ) effective Hamiltonian if the electron Rabi field is parallel (antiparallel) to the spin-locked electron spin. Then, similar to the treatment in CP, the initial density operator $\hat{\rho}(0) = \hat{S}_z$ evolves under either the ZQ or DQ effective Hamiltonian to become:

$$\hat{\rho}^{ZQ}(t) = \hat{S}_z \cos^2(\omega_{\text{NOVEL}}t) + \hat{I}_z \sin^2(\omega_{\text{NOVEL}}t) + \sin(2\omega_{\text{NOVEL}}t)(\hat{S}_y\hat{I}_x - \hat{S}_x\hat{I}_y), \tag{14}$$

$$\hat{\rho}^{DQ}(t) = \hat{S}_z \cos^2(\omega_{\text{NOVEL}}t) - \hat{I}_z \sin^2(\omega_{\text{NOVEL}}t) + \sin(2\omega_{\text{NOVEL}}t)(\hat{S}_y\hat{I}_x + \hat{S}_x\hat{I}_y), \tag{15}$$

where $\omega_{\text{NOVEL}} = B_{zx}/4$ denotes the NOVEL build-up rate, and that the DQ transfer has an opposite sign relative to the ZQ case. Although the $\sin^2(\omega_{\text{NOVEL}}t)$ term (Eq. 14) implies transient oscillations as in CP, such effects are not easily observed in experiments. This is because most state-of-the-art

DNP experiments observe only bulk $^1$H instead of an isolated $^1$H spin, and nuclear spin diffusion dampens such transients. Although there have been a few reports of such transient-like features in literature, they are mostly performed on single crystals, or the experiments were not optimized for such purposes.[25–28] Encouraged by these early findings, we envision that our theoretical framework here could motivate some new experimental efforts in observing these transients at high fields, realizing electron-nucleus distance measurement using DNP.

### 2.3 Solid effect (SE)

There are several theoretical approaches that can analyze the SE in literature.[12,13,29,30] One could apply perturbation theory to determine the degree of state mixing between the Zeeman eigenstates, and this method is best suited if the electron nutation frequency, $\omega_{1S}$, is small relative to the electron relaxation rate, $T_{1e}^{-1}$. Here, we will show an approach similar to the framework shown by Jain et. al,[27] where AHT and the product-operator formalism were used. We begin with a similar Hamiltonian as in Eq. 8 except for an inclusion of an electron offset term, $\Omega \hat{S}_z$:

$$\hat{\mathcal{H}}_{SE} = \Omega \hat{S}_z - \omega_{0I}\hat{I}_z + A_{zz}\hat{S}_z\hat{I}_z + A_{zx}\hat{S}_z\hat{I}_x + A_{zy}\hat{S}_z\hat{I}_y + \omega_{1S}\hat{S}_x, \qquad (16)$$

Additionally, we show the matrix representation of Hamiltonian (Eq. 16) using the Zeeman eigenbases $|\alpha\alpha\rangle$, $|\alpha\beta\rangle$, $|\beta\alpha\rangle$, and $|\beta\beta\rangle$ for an electron-nucleus system $|en\rangle$:

$$\hat{\mathcal{H}}_{SE} \equiv \frac{1}{4}\begin{bmatrix} A_{zz} - 2\omega_{0I} + 2\Omega & A_{zx} - iA_{zy} & 2\omega_{1s} & 0 \\ A_{zx} - iA_{zy} & -A_{zz} + 2\omega_{0I} + 2\Omega & 0 & 2\omega_{1s} \\ 2\omega_{1s} & 0 & -A_{zz} - 2\omega_{0I} - 2\Omega & -A_{zx} + iA_{zy} \\ 0 & 2\omega_{1s} & -A_{zx} - iA_{zy} & A_{zz} + 2\omega_{0I} - 2\Omega \end{bmatrix} \qquad (17)$$

Here, we emphasize the importance of recognizing the SE matrix representation because it will be used later to derive the CE DNP matching condition. Next, we simplify the Hamiltonian by introducing the pseudo-secular term $B_{zx}\hat{S}_z\hat{I}_x$ (see Eq. 9), and apply a propagator $\hat{U}_t = \exp(-i\theta\hat{S}_y)$ with $\theta = \tan^{-1}(\omega_{1S}/\Omega)$, so that the electron z-axis is aligned along the effective field $\omega_{\text{eff}} = \sqrt{\omega_{1s}^2 + \Omega^2}$:[3]

---

[3] Note that the initial density operator is now $\rho(0) = \cos\theta\,\hat{S}_z - \sin\theta\,\hat{S}_x$.

$$\widehat{\mathcal{H}}_{\text{SE}} = \omega_{\text{eff}} \cos \theta \, \hat{S}_z - \omega_{0I} \hat{I}_z + A_{zz} \hat{S}_z \hat{I}_z + B_{zx} \hat{S}_z \hat{I}_x + \omega_{\text{eff}} \sin \theta \, \hat{S}_x$$

(18)

$$\widehat{\mathcal{H}}_t = \widehat{U}_t^{-1} \widehat{\mathcal{H}}_{\text{SE}} \widehat{U}_t = \omega_{\text{eff}} \hat{S}_z - \omega_{0I} \hat{I}_z + \left(A_{zz} \hat{I}_z + B_{zx} \hat{I}_x\right)\left(\cos \theta \, \hat{S}_z - \sin \theta \, \hat{S}_x\right),$$

followed by another interaction-frame transformation using $\widehat{U}_1 = \exp\bigl(-i(\omega_{\text{eff}} \hat{S}_z - \omega_{0I} \hat{I}_z)t\bigr)$:

$$\begin{aligned}
\widehat{\widehat{\mathcal{H}}}(t) &= \widehat{U}_1^{-1}(t) \widehat{\mathcal{H}}_t \widehat{U}_1(t) - \left(\omega_{\text{eff}} \hat{S}_z - \omega_{0I} \hat{I}_z\right) \\
&= \left(A_{zz} \hat{I}_z + B_{zx} (\hat{I}_x \cos \omega_{0I} t + \hat{I}_y \sin \omega_{0I} t)\right) \left(\cos \theta \, \hat{S}_z - \sin \theta \, (\hat{S}_x \cos \omega_{\text{eff}} t - \hat{S}_y \sin \omega_{\text{eff}} t)\right).
\end{aligned}$$

(19)

Similarly, the Hamiltonian $\widehat{\widehat{\mathcal{H}}}(t)$ becomes time-independent if $\omega_{\text{eff}} = \pm \omega_{0I}$ (see Appendix 1) and AHT is applied with the condition $\omega_{0I} \gg A_{zz} \sin \theta, B_{zx} \cos \theta$:

$$\begin{aligned}
\widehat{\widehat{\mathcal{H}}}(t) &= A_{zz} \cos \theta \, \hat{S}_z \hat{I}_z \\
&\quad - \frac{B_{zx} \sin \theta}{2} \left[(\hat{S}_x \hat{I}_x \mp \hat{S}_y \hat{I}_y) + (\hat{S}_x \hat{I}_x \pm \hat{S}_y \hat{I}_y) \cos 2\omega_{0I} t + (\hat{S}_x \hat{I}_y \mp \hat{S}_y \hat{I}_x) \sin 2\omega_{0I} t\right] \\
&\quad - A_{zz} \sin \theta \, (\hat{S}_x \hat{I}_z \cos \omega_{0I} t \mp \hat{S}_y \hat{I}_z \sin \omega_{0I} t) \\
&\quad + B_{zx} \cos \theta \, (\hat{S}_z \hat{I}_x \cos \omega_{0I} t + \hat{S}_z \hat{I}_y \sin \omega_{0I} t).
\end{aligned}$$

(20)

$$\overline{\widehat{\widehat{\mathcal{H}}}} = \frac{\omega_{0I}}{2\pi} \int_0^{2\pi/\omega_{0I}} \widehat{\widehat{\mathcal{H}}}(t) \, dt = \begin{cases} -\dfrac{B_{zx} \sin \theta}{2} (\hat{S}_x \hat{I}_x + \hat{S}_y \hat{I}_y) + A_{zz} \hat{S}_z \hat{I}_z \cos \theta & \text{if } \omega_{\text{eff}} = -\omega_{0I} \\ -\dfrac{B_{zx} \sin \theta}{2} (\hat{S}_x \hat{I}_x - \hat{S}_y \hat{I}_y) + A_{zz} \hat{S}_z \hat{I}_z \cos \theta & \text{if } \omega_{\text{eff}} = +\omega_{0I} \end{cases},$$

(21)

where ZQ and DQ fictitious spin-1/2 operators are again obtained (see Eqs 5 and 13). Although there is now an extra $A_{zz} \hat{S}_z \hat{I}_z$ term in Eq. 21 compared to the CP and NOVEL cases, its effect can be safely ignored because the $\hat{S}_z \hat{I}_z$ operator represents an identity operator in the ZQ/DQ subspaces; it commutes with all operators. This is evident by inspecting the matrix representation of Eq. 21 for the case of $\omega_{\text{eff}} = -\omega_{0I}$ (ZQ condition):

$$\overline{\widehat{\widehat{\mathcal{H}}}}_{\text{ZQ}} \equiv \frac{1}{4} \begin{bmatrix} A_{zz} \cos \theta & 0 & 0 & 0 \\ 0 & -A_{zz} \cos \theta & -B_{zx} \sin \theta & 0 \\ 0 & -B_{zx} \sin \theta & -A_{zz} \cos \theta & 0 \\ 0 & 0 & 0 & A_{zz} \cos \theta \end{bmatrix}.$$

(22)

Following from Eq. 21, the initial density operator along the effective field, $\rho(0) = \hat{S}_z$ will evolve under the ZQ/DQ Hamiltonian to become:

$$\hat{\rho}^{ZQ}(t) = \hat{S}_z \cos^2(\omega_{SE}t) + \hat{I}_z \sin^2(\omega_{SE}t) + (\hat{S}_y\hat{I}_x - \hat{S}_x\hat{I}_y)\sin(2\omega_{SE}t)$$

$$\hat{\rho}^{DQ}(t) = \hat{S}_z \cos^2(\omega_{SE}t) - \hat{I}_z \sin^2(\omega_{SE}t) + (\hat{S}_y\hat{I}_x + \hat{S}_x\hat{I}_y)\sin(2\omega_{SE}t)$$

(23)

where the SE build-up rate is $\omega_{SE} = B_{zx}\sin\theta/4$. Since the results were obtained using a generalized expression with minimal assumptions, the weak μw irradiation case ($\omega_{1s} \ll \omega_{0I}$) should converge to the same results shown in the literature.[13] Hence, the matching conditions are:

$$\omega_{eff} = \sqrt{\omega_{1s}^2 + \Omega^2} = \pm\omega_{0I} \text{ for general cases}$$

$$\Rightarrow \Omega \sim \pm\omega_{0I} \text{ for weak μw cases,}$$

(24)

and the initial nuclei polarization build-up (from Eq. 23) is given by $\langle\hat{\rho}(t \ll 1)|\hat{I}_z\rangle \propto \sin^2(\omega_{SE}t) \sim \omega_{SE}^2 t^2$, where $\omega_{SE}^2 = (B_{zx}\omega_{1s}/4\omega_{0I})^2$. We have derived the well-known SE matching conditions $\Omega \sim \pm\omega_{0I}$, and the $\hat{I}_z(t \ll 1) \propto \omega_{0I}^{-2}$ dependence yields the well-known observation that the enhancement factor scales by a factor of $\sim\omega_{0I}^{-2}$—which is true if the Rabi field $\omega_{1s}$, relaxation rates, and all other factors remain constant when the $B_0$ field increases.[30]

### 2.4 Cross effect (CE)

The CE was first discovered when the DNP field profile showed changes when higher radical concentrations were used.[10,11] The effect was then exploited by tethering two monomeric nitroxide radicals to form a biradical. [31–33] The underlying CE mechanism was explained theoretically using perturbation theory for the static and the MAS cases.[9,30–34] In the theoretical analyses, the perturbation treatment was applied twice because there is no off-diagonal term that directly connects the two degenerate energy eigenstates. Such successive perturbative treatment might lose important insights because some terms are discarded in each perturbative step. We will now revisit the CE theory in the static case and demonstrate that new insights are obtained using the unified theory.

A generic lab-frame Hamiltonian for a two-electron-one-nucleus system is given by:

$$\hat{\mathcal{H}} = d(3\hat{S}_{1z}\hat{S}_{2z} - \hat{S}_1 \cdot \hat{S}_2) - 2J\hat{S}_1 \cdot \hat{S}_2 + A_{zz}^{(1)}\hat{S}_{1z}\hat{I}_z + A_{zx}^{(1)}\hat{S}_{1z}\hat{I}_x + A_{zy}^{(1)}\hat{S}_{1z}\hat{I}_y \ldots$$

(25)

$$+A_{zz}^{(2)}\hat{S}_{2z}\hat{I}_{z} + A_{zx}^{(2)}\hat{S}_{2z}\hat{I}_{x} + A_{zy}^{(2)}\hat{S}_{2z}\hat{I}_{y} + \omega_{0S1}\hat{S}_{1z} + \omega_{0S2}\hat{S}_{2z} - \omega_{0I}\hat{I}_{z},$$

where $d$ and $J$ represent the dipolar coupling and the exchange interaction between the two electrons, respectively. Note that the $A_{zz}^{(n)}$ component is secular, while the $A_{zx}^{(n)}$ and $A_{zy}^{(n)}$ components are not. The index $n$ labels the hyperfine interaction between the electron and the first or second nucleus ($n$ =1 or 2). $\omega_{0I(S)}$ denotes the Larmor frequency of the nucleus (electrons). Note that the usual tilted-frame transformation along $\hat{I}_z$ to remove $\hat{S}_z\hat{I}_y$ cannot be done here for both electron-nuclei spin pairs unless one assumes that the sizes and signs of $A_{zy}^{(n)}/A_{zx}^{(n)}$ are the same for both electron-nucleus pairs ($n$ =1 or 2).

Fig. 2 Matrix representation of the Hamiltonian that describes the two-electron-one-nucleus spin system, where the CE subspace is comprised of the middle 4x4 Eigenbases $\{|\alpha_e\beta_e\alpha_n\rangle, |\alpha_e\beta_e\beta_n\rangle, |\beta_e\alpha_e\alpha_n\rangle, |\beta_e\alpha_e\beta_n\rangle\}$. Note that $\Sigma\omega_\pm = 2(\omega_{0S1} + \omega_{0S2} \pm \omega_{0I})$ and $\Delta\omega_\pm = (\omega_{0S1} - \omega_{0S2} \pm \omega_{0I})$. The unfilled matrix elements are zero.

By inspecting the matrix representation of the Hamiltonian (Eq. 25) excluding the µw field, it is clear that the full Hamiltonian is block diagonalized with one central 4×4 block (Fig. 2) and two 2×2 blocks. We call the central 4×4 block the CE subspace, which will be our focus because the other two blocks are relevant only for NMR transitions. The matrix representation of the CE subspace can be rewritten in a more compact form to become:

$$\hat{\mathcal{H}}^{\mathrm{CE}} \equiv \frac{J-d}{2}\hat{\mathbb{E}} \tag{26}$$

$$+\frac{1}{4}\begin{bmatrix} \Delta A_{zz} - 2\omega_{0I} + 2\Delta\omega & \Delta A_{zx} - i\Delta A_{zy} & -2(2J+d) & 0 \\ \Delta A_{zx} + i\Delta A_{zy} & -\Delta A_{zz} + 2\omega_{0I} + 2\Delta\omega & 0 & -2(2J+d) \\ -2(2J+d) & 0 & -\Delta A_{zz} - 2\omega_{0I} - 2\Delta\omega & -\Delta A_{zx} + i\Delta A_{zy} \\ 0 & -2(2J+d) & -\Delta A_{zx} - i\Delta A_{zy} & \Delta A_{zz} + 2\omega_{0I} - 2\Delta\omega \end{bmatrix}$$

where $\Delta\omega = \omega_{0S1} - \omega_{0S2}$, $\Delta A_{zz} = A_{zz}^{(1)} - A_{zz}^{(2)}$, $\Delta A_{zx} = A_{zx}^{(1)} - A_{zx}^{(2)}$, $\Delta A_{zy} = A_{zy}^{(1)} - A_{zy}^{(2)}$, and $\hat{\mathbb{E}}$ is the neglectable identity operator.[4] At this point, we can directly quote the matching conditions and the effective Hamiltonian for the CE because the $\hat{\mathcal{H}}^{\text{CE}}$ matrix (Eq. 26) is mathematically identical to the $\hat{\mathcal{H}}_{\text{SE}}$ matrix (Eq. 17) except for the definition of the symbols. For instance, $\Omega$, $\omega_{1S}$, $A_{zz}$, $A_{zx}$, and $A_{zy}$ in SE are now analogous to $\Delta\omega$, $-(2J+d)$, $\Delta A_{zz}$, $\Delta A_{zx}$, and $\Delta A_{zy}$, respectively in CE. Thus, we can define new fictitious spin-1/2 operators for this CE subspace and rewrite Eq. 27:

$$\hat{\mathcal{H}}^{\text{CE}} = \frac{J-d}{2}\hat{\mathbb{E}} + \Delta\omega\hat{S}_z^{\text{CE}} - \omega_{0I}\hat{I}_z^{\text{CE}} + \Delta A_{zz}\hat{S}_z^{\text{CE}}\hat{I}_z^{\text{CE}} + \Delta A_{zx}\hat{S}_z^{\text{CE}}\hat{I}_x^{\text{CE}} + \Delta A_{zy}\hat{S}_z^{\text{CE}}\hat{I}_y^{\text{CE}} - (2J+d)\hat{S}_x^{\text{CE}}, \qquad (27)$$

where the eigenstates of the fictitious operator $\hat{S}_z^{\text{CE}}$ are $|\alpha_e\beta_e\rangle$ and $|\beta_e\alpha_e\rangle$. By ensuring that the AHT assumption made in Eq. 21 remains valid here, the SE results and matching conditions can be directly adapted for the CE case:

$$\omega_{0I} = \pm\sqrt{(2J+d)^2 + \Delta\omega^2}$$
$$\sim \pm\Delta\omega \text{ if } |2J+d| \ll |\omega_{0I}|. \qquad (28)$$

Similarly, the build-up rate is

$$\omega_{\text{CE}} = \frac{\Delta B_{zx}(2J+d)}{4\omega_{0I}}, \qquad (29)$$

where $\Delta B_{zx} = \sqrt{\Delta A_{zx}^2 + \Delta A_{zy}^2}$, which can be regarded as the Pythagorean sum of the size differences between the two pseudo-secular hyperfine interactions. We call this term the differential hyperfine interaction. Then, by directly adapting the results from SE (Eq. 23), one can express that the polarization is transferred from $\hat{S}_z^{\text{CE}}$ to $\hat{I}_z^{\text{CE}}$ via the fictitious operator $\hat{S}_x^{\Delta,\text{CE}}$ in the CE-ZQ double subspace, i.e., $\hat{S}_z^{CE} \xrightarrow{2\omega_{CE}\hat{S}_x^{\Delta,CE}} \hat{I}_z^{CE}$ (see Eq. 7). The CE transfer mechanism has been analytically derived with minimal efforts and assumptions.

---

[4] As discussed earlier, identity operator commutes with all operators, and, hence, negligible.

Note that the buildup rate $\omega_{CE}$ (Eq. 29) derived here is similar to those shown in the literature, [12,34–36] except on the significance of $\Delta B_{zx}$ in mediating DNP, and that the CE buildup curve should also exhibit a familiar transient oscillation just like other polarization-transfer sequences discussed earlier. We emphasize that the description refers only to the static case. For MAS, where the energy levels are crossings, and the matching conditions are adiabatically swept, one would expect an exponential buildup behaviour—as correctly predicted by the level anti-crossing (LAC) framework and Landau-Zener equation. [12,39]

Although the source of polarization $\hat{S}_z^{CE}$ is negligible at thermal equilibrium, it can be enlarged by saturating either electron. For example, if the first electron is saturated via microwaves at $\omega_{\mu w} = \omega_{0S1}$ (exciting both $|\beta_e\beta_e\rangle \leftrightarrow |\alpha_e\beta_e\rangle$ and $|\beta_e\alpha_e\rangle \leftrightarrow |\alpha_e\alpha_e\rangle$ transitions), this will indirectly create a population difference between the $|\alpha_e\beta_e\rangle$ and $|\beta_e\alpha_e\rangle$ states—this prepares a non-zero $\hat{S}_z^{CE}$. Again, we emphasize that it is the *difference* between the two electron spin polarization, rather than the absolute electron polarization, that is responsible for mediating CE DNP. Hence, we expect CE DNP to be more efficient if the two electrons have very different $T_1$ values,[40,41] with the slower-relaxing electron being saturated prior to the CE matching condition (Eq. 28). This is to say that the CE mechanism is a two-step process where the first step requires the two-electron polarization difference resulting from μw to saturate an electron or other approaches, and the second step involves a passive three-spin flip process that does not require active perturbations including μw irradiation. Thus, these two processes do not need to occur simultaneously, and this unique feature was exemplified in the MAS case, where the saturation and polarization steps happen at different rotor angles. Following this idea, a *Gedankenexperiment* where CE DNP can be mediated by generating polarization difference between two electrons with selective optical pumping (without microwave irradiation) was proposed recently.[42] Besides CE, we also note that the fictitious operator formalism on a three-spin system was also applied to understanding chemically induced DNP.[43]

In summary, we realize that the matrix representation of the CE Hamiltonian is mathematically similar to the SE Hamiltonian. This allows us to directly adopt the SE results for CE with minimal approximations. We will later show how the position of the ¹H nuclei in a biradical and the differential hyperfine interaction $\Delta B_{zx}$ could affect the CE DNP performance (section 3.3).

| Table 1. Unified theoretical framework for CP, NOVEL, SE, and CE | | | | |
|---|---|---|---|---|
|  | CP | NOVEL | SE | CE |
| Initial Hamiltonians in respective frames | Equation 1 | Equation 8 | Equation 16 | Equation 27 |
| Effective Hamiltonian | $\widehat{\overline{\mathcal{H}}} = d_{IS}(\hat{S}_x\hat{I}_x + \hat{S}_y\hat{I}_y)$ | $\widehat{\overline{\mathcal{H}}} = -\frac{B_{zx}}{2}(\hat{S}_x\hat{I}_x \pm \hat{S}_y\hat{I}_y)$ | $\widehat{\overline{\mathcal{H}}} = -\frac{B_{zx}\omega_{1S}}{2\omega_{0I}}(\hat{S}_x\hat{I}_x \pm \hat{S}_y\hat{I}_y)$ $+A_{zz}\hat{S}_z\hat{I}_z\cos\theta$ | $\widehat{\overline{\mathcal{H}}} = \frac{\Delta B_{zx}(2J+d)}{2\omega_{0I}}(\hat{S}_x\hat{I}_x \pm \hat{S}_y\hat{I}_y)$ $+\Delta A_{zz}\hat{S}_z\hat{I}_z\cos\theta$ |
| AHT assumptions | $\omega_{1I}, \omega_{1S} > d_{IS}$ | $\omega_{1S}, \omega_{0I} > B_{zx}$ | $\omega_{0I}, \Omega > A_{zz}\cos\theta, B_{zx}\sin\theta$ | $\omega_{0I}, \Delta\omega > \Delta A_{zz}\cos\theta, \Delta B_{zx}\sin\theta$ |
| Initial density operator $\hat{\rho}(0)$ | $\hat{I}_z$ | $\hat{S}_z$ | $\hat{S}_z$ | $\hat{S}_z^{CE}$ |
| Matching condition | $\omega_{1S} = \omega_{1I}$ | $\omega_{1S} = \omega_{0I}$ | $\omega_{0I} = \pm\omega_{eff} = \pm\sqrt{\Omega^2 + \omega_{1S}^2}$ | $\omega_{0I} = \pm\omega_{eff} = \pm\sqrt{\Delta\omega^2 + (2J+d)^2}$ |
| Build-up rate | $\omega_{CP} = \frac{d_{IS}}{2}$ | $\omega_{NOVEL} = \frac{B_{zx}}{4}$ | $\omega_{SE} = \frac{B_{zx}\omega_{1S}}{4\omega_{0I}}$ | $\omega_{CE} = \frac{\Delta B_{zx}(2J+d)}{4\omega_{0I}}$ |

## 3 Results and discussion

We have analyzed and concluded that CP, NOVEL, SE, and CE could be described using a unified theoretical framework, yielding the effective Hamiltonians, matching conditions, build-up rates, etc. (Table 1). Next, we will analyze the situation in which the matching conditions are not exactly fulfilled, i.e., a slight mismatch is present.

### 3.1 Treating mismatch in DNP matching condition

Note that we have only derived the effective Hamiltonians and buildup rates for which the DNP matching conditions are perfectly satisfied. In practical situations, each electron spin packet might experience a different effective field $\omega_{eff}(\Omega, \omega_{1S})$ due to μw field ($\omega_{1S}$) inhomogeneity or *g*-anisotropy/offset frequencies ($\Omega$). Consequently, only a fraction of electron spins fulfils the matching conditions, and the remaining spins experience mismatches in varying magnitudes, which will be analyzed here. First, we will consider the SE case in which a mismatch frequency $\delta_{SE} \neq 0$ is present:

$$\delta_{SE} = \omega_{eff} + \omega_{0I} \text{ (ZQ) or } \delta_{SE} = \omega_{eff} - \omega_{0I} \text{ (DQ).} \quad (30)$$

Then, we perform an interaction-frame transformation on Eq. 18 with the propagator $\hat{U}_{1'} = \exp(-i(\omega_{eff}\hat{S}_z - \omega_{0I}\hat{I}_z - \delta_{SE}\hat{S}_z)t)$:

$$\begin{aligned}
\widehat{\widetilde{\mathcal{H}}}' &= \hat{U}_{1'}^{-1}(t)\hat{\mathcal{H}}_t\hat{U}_{1'}(t) - \omega_{eff}\hat{S}_z + \omega_{0I}\hat{I}_z + \delta_{SE}\hat{S}_z^\Sigma + \delta_{SE}\hat{S}_z^\Delta \\
&= A_{zz}\cos\theta\,\hat{S}_z\hat{I}_z + \delta_{SE}\hat{S}_z^\Sigma + \delta_{SE}\hat{S}_z^\Delta \\
&\quad - \frac{B_{zx}\sin\theta}{2}\big[(\hat{S}_x\hat{I}_x - \hat{S}_y\hat{I}_y)\cos((\omega_{eff} - \delta_{SE} - \omega_{0I})t) \\
&\qquad\qquad\qquad - (\hat{S}_x\hat{I}_y + \hat{S}_y\hat{I}_x)\sin((\omega_{eff} - \delta_{SE} - \omega_{0I})t)\big] \\
&\quad - \frac{B_{zx}\sin\theta}{2}\big[(\hat{S}_x\hat{I}_x + \hat{S}_y\hat{I}_y)\cos((\omega_{eff} - \delta_{SE} + \omega_{0I})t) \\
&\qquad\qquad\qquad + (\hat{S}_x\hat{I}_y - \hat{S}_y\hat{I}_x)\sin((\omega_{eff} - \delta_{SE} + \omega_{0I})t)\big] \\
&\quad - A_{zz}\sin\theta\,(\hat{S}_x\hat{I}_z\cos(\omega_{eff} - \delta_{SE})t - \hat{S}_y\hat{I}_z\sin(\omega_{eff} - \delta_{SE})t) \\
&\quad + B_{zx}\cos\theta\,(\hat{S}_z\hat{I}_x\cos\omega_{0I}t + \hat{S}_z\hat{I}_y\sin\omega_{0I}t),
\end{aligned} \quad (31)$$

where we have used the fictitious operator $\hat{S}_z^\Sigma = (\hat{S}_z + \hat{I}_z)/2$ and $\hat{S}_z^\Delta = (\hat{S}_z - \hat{I}_z)/2$ (see Appendix 3). If $A_{zz}, B_{zx} \ll \omega_{0I}$, we can apply AHT and obtain:

$$\begin{aligned}
\widehat{\overline{\mathcal{H}}}' &= \frac{\omega_{eff} - \delta_{SE}}{2\pi}\int_0^{2\pi/(\omega_{eff}-\delta_{SE})} \widehat{\widetilde{\mathcal{H}}}'(t)\,dt \\
&= \begin{cases} -\frac{B_{zx}\sin\theta}{2}(\hat{S}_x\hat{I}_x + \hat{S}_y\hat{I}_y) + \delta_{SE}\hat{S}_z^\Sigma + \delta_{SE}\hat{S}_z^\Delta + A_{zz}\cos\theta\,\hat{S}_z\hat{I}_z & \text{if } \omega_{eff} - \delta_{SE} + \omega_{0I} = 0 \\ -\frac{B_{zx}\sin\theta}{2}(\hat{S}_x\hat{I}_x - \hat{S}_y\hat{I}_y) + \delta_{SE}\hat{S}_z^\Sigma + \delta_{SE}\hat{S}_z^\Delta + A_{zz}\cos\theta\,\hat{S}_z\hat{I}_z & \text{if } \omega_{eff} - \delta_{SE} - \omega_{0I} = 0 \end{cases} \\
&= \begin{cases} -\frac{B_{zx}\sin\theta}{2}\hat{S}_x^\Delta + \delta_{SE}\hat{S}_z^\Sigma + \delta_{SE}\hat{S}_z^\Delta + A_{zz}\cos\theta\,\hat{S}_z\hat{I}_z & \text{if } \omega_{eff} - \delta_{SE} + \omega_{0I} = 0 \\ -\frac{B_{zx}\sin\theta}{2}\hat{S}_x^\Sigma + \delta_{SE}\hat{S}_z^\Sigma + \delta_{SE}\hat{S}_z^\Delta + A_{zz}\cos\theta\,\hat{S}_z\hat{I}_z & \text{if } \omega_{eff} - \delta_{SE} - \omega_{0I} = 0 \end{cases}.
\end{aligned} \quad (32)$$

Then, the evolution of $\rho'(t)$ can be computed using the LvN equation with $\rho(0) = \hat{S}_z = \hat{S}_z^\Sigma + \hat{S}_z^\Delta$:

$$\hat{\rho}'(t) = \begin{cases} \hat{S}_z(1 - \cos^2\zeta_{SE}\sin^2\omega'_{SE}t) + \hat{I}_z\cos^2\zeta_{SE}\sin^2\omega'_{SE}t - \hat{S}_x^\Delta\sin 2\zeta_{SE}\sin^2\omega'_{SE}t + \hat{S}_y^\Delta\cos\zeta_{SE}\sin 2\omega'_{SE}t \\ \hat{S}_z(1 - \cos^2\zeta_{SE}\sin^2\omega'_{SE}t) - \hat{I}_z\cos^2\zeta_{SE}\sin^2\omega'_{SE}t - \hat{S}_x^\Sigma\sin 2\zeta_{SE}\sin^2\omega'_{SE}t + \hat{S}_y^\Sigma\cos\zeta_{SE}\sin 2\omega'_{SE}t \end{cases}, \quad (33)$$

where $\zeta_{SE} = \tan^{-1}(\delta_{SE}/(2\omega_{SE}))$,

$$\omega'_{SE} = \sqrt{\omega_{SE}^2(\Omega) + (\delta_{SE}/2)^2} = \sqrt{\frac{B_{zx}^2 \omega_{1S}^2}{16(\omega_{1S}^2 + \Omega^2)} + \frac{\delta_{SE}^2}{4}}, \quad (34)$$

and $\omega_{SE}(\Omega) = B_{zx}\omega_{1S}/(4\omega_{eff}(\Omega))$. Note that the new build-up rates $\omega'_{SE}$ is apparently faster when a mismatch is present, albeit that the maximum polarization is lower by a factor of $\cos^2 \zeta_{SE}$. This phenomenon is mathematically analogous to the situation in which an off-resonance inversion $\pi$ pulse was applied to a spin, i.e., the polarization cannot be fully inverted despite experiencing a larger effective field. The mismatch is now treated as offsets in the ZQ and DQ fictitious spin-1/2 subspaces. Similarly, we can extend the theorem for the CE case by introducing mismatch $\delta_{CE} = \omega_{eff,CE} \pm \omega_{0I}$, leading to $\omega'_{CE} = \sqrt{\omega_{CE}^2(\Delta B_{zx}, d) + (\delta_{CE}/2)^2}$ and $\zeta_{CE} = \tan^{-1}(\delta_{CE}/(2\omega_{CE}))$ (vide infra, see Fig. 7c).

Table 2. Parameters used in theoretical analyses and numerical simulations. The PAS-to-lab frame Euler angles are $(60°, 45°, 36°)$.

|  | NOVEL | SE | CE |
|---|---|---|---|
| **Magnetic field $B_0$** | 0.35 T | 5 T | 5 T |
| **g factor** | 2.003 | 2.003 | $g_{e1}$=2.003000<br>$g_{e2}$ is 2.006041 |
| **Coordinate (PAS)** | $^1$H (0, 0, 0)<br>e (0, 0, $r_{eH}$) | $^1$H (0, 0, 0)<br>e (0, 0, $r_{eH}$) | $^1$H (0, 0, 0)<br>e1 (0, 0, $r_{e1H}$)<br>e2 $0.8 r_{e1H}$ ($\sin 135°$, 0, $\cos 135°$) |
| **Microwave Rabi Field $\omega_{1S}$** | $\omega_{0I}$ | 4 MHz | - |
| **$\hat{\rho}(0)$** | $\hat{S}_{1z}$ | $\hat{S}_{1z}$ | Saturated: $\hat{S}_{1z}$<br>Inverted: $\hat{S}_{1z} - \hat{S}_{2z}$ |

### 3.2 Theoretical results and numerical simulations

To verify the theoretical results, we compare the evolution of density operators using the effective Hamiltonians (Table 1) with numerically simulated results from *Spinach*.[44] For easier analysis, relaxation effects and *g*-anisotropy are not included at this stage, and other parameters are listed in Table 2. The field strengths for the SE and CE were chosen to be 5 T, but we restricted NOVEL to 0.35 T because it is not yet feasible to perform this experiment at higher fields. Note that the μw irradiation is employed throughout the sequence for NOVEL and SE. For CE, ideal μw pulses were applied only in the beginning to prepare the electron polarization difference and subsequently 'turned off'.

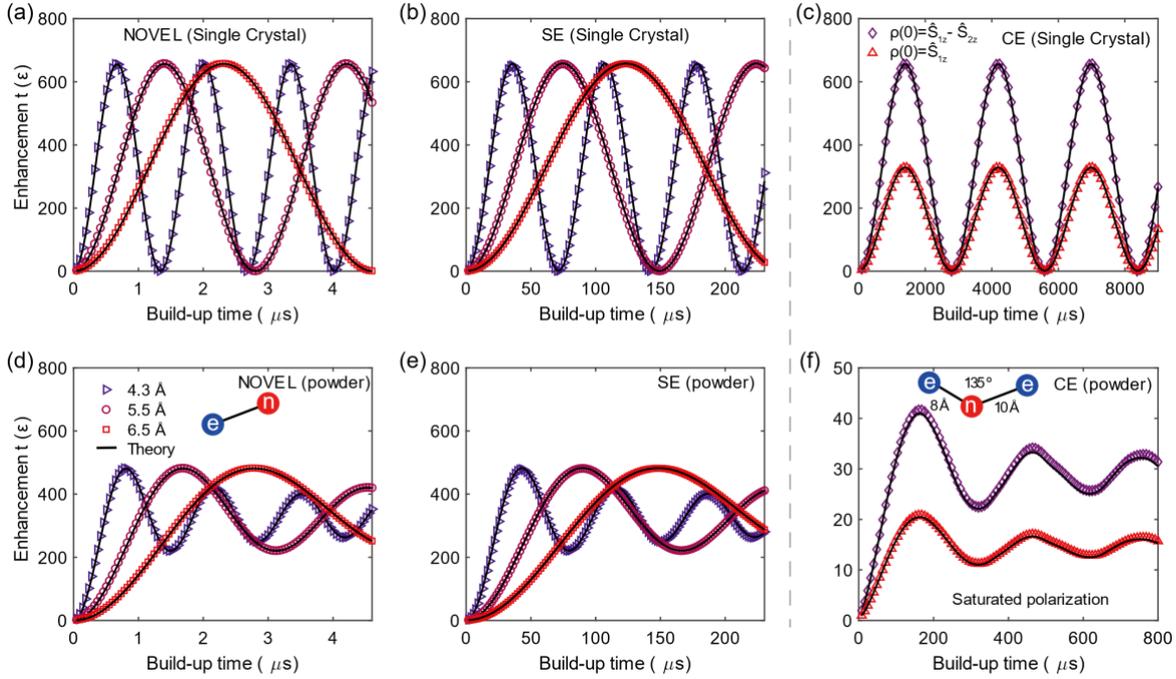

Fig. 3 Plots of $^1$H enhancement calculated by theory (line) and numerical simulations for (a, d) NOVEL, (b, e) SE, and (c, f) CE on (a-c) single-crystal or (d-f) powdered samples. DNP with different e-$^1$H distances (4.3 Å, 5.5 Å, and 6.5 Å are examined in NOVEL and SE. Two different initial states are considered in CE (c and f): saturated second electron spin $\rho(0) = \hat{S}_{1z}$ (red) and inverted second electron spin, $\rho(0) = \hat{S}_{1z} - \hat{S}_{2z} = \hat{S}_z^{CE}$ (violet). The two e-$^1$H distances are 10 Å and 8 Å, respectively. The e-$^1$H angle is 135°. The powder averages were performed using the two-angle Lebedev grids with rank 131 provided in Spinach.

The results of numerical simulations agree with the theoretical predictions exceptionally well (Fig. 3) in all three DNP cases. In particular, the theoretically derived DNP buildup rates ($\omega_{\text{NOVEL}} = B_{zx}/4$, $\omega_{\text{SE}} = B_{zx}\omega_{1S}/4\omega_{0I}$, and $\omega_{\text{CE}} = \Delta B_{zx}(2J+d)/4\omega_{0I}$) for different electron-nucleus distances are verified to be correct. Note that a near ~100% ($\varepsilon$~658) polarization transfer is possible for single crystals in NOVEL (Fig. 3a), but the SE (Fig. 3b) has a marginally lower enhancement due to a small projection loss between $\hat{S}_{1z}$ and the direction of the effective field $\omega_{\text{eff}}$. For CE (Fig. 3c), only ~50 % transfer ($\varepsilon$~329) can be obtained if the second electron is saturated—as to how it would normally be performed in CW DNP ($\rho(0) = \hat{S}_{1z}$ only). However, if the polarization of the second electron is inverted with a $\pi$ pulse using a pulsed microwave source, i.e., to prepare an initial state of $\rho(0) =$

$\hat{S}_{1z} - \hat{S}_{2z} = \hat{S}_z^{CE}$, a ~ 100 % transfer can be achieved again. Thus, we demonstrated that the fictitious operator ($\hat{S}_z^{CE}$) used in our unified theory has shown a new and intuitive insight in this *pulsed* cross effect experiment.[45]

For powdered samples, the maximum transient polarization is ε~482 (Fig. 3d-e) in SE and NOVEL, which corresponds to a transfer efficiency of 482/658~73 %—a known benchmark value obtainable by γ-encoded sequences including CP.[46–49] This can be inferred from Eq. 13 and Eq. 21, where the Hamiltonians (or $B_{zx}$) are γ-independent.[5] Again, it is clear that some developed concepts that existed in the well-familiarized conventional ssNMR techniques can be directly adapted for DNP cases, and perhaps shed new lights on analyzing the existing DNP sequence. For instance, one can evaluate the robustness of a pulse sequence by inspecting if the effective Hamiltonian of a pulse sequence is γ-encoded. Similarly, we envision such an evaluation strategy can be better exploited when developing new DNP sequences, especially when the high-frequency pulsed microwave technology becomes available in the future.[19] The CE performance on powdered samples (Fig. 3f) is much weaker than SE's and NOVEL's because only a small fraction of the crystallites satisfies the orientation-dependent $d(\alpha, \beta, \gamma)$ in CE matching condition (Eq. 28).

Next, *g*-anisotropy is included in the simulations to resemble actual DNP experiments. The buildup curves for several crystallites with different SE matching conditions and three different e-[1]H spin systems (Fig. 4a) were examined. All three systems have the same spin interactions except the e-[1]H Euler angles relative to the *g* tensor. The simulated curves fit the calculated results from the unified theory well—if the mismatched situations are also considered (Eq. 33). In other words, the theoretically calculated curves (Fig. 4b-d) were not performed using a single crystal, but an entire powder spectrum that includes crystallites that do not exactly satisfy the matching conditions.[6] Besides, it is evident that the buildup profiles are sensitive to the e-[1]H Euler angles, which imply that it is theoretically possible to determine the full e-$^1$H dipolar coupling tensor from DNP. If experimentally proven, the technique could have important applications on paramagnetic biomolecules or materials.

---

[5] Note that only $A_{zx}$ and $A_{zy}$ depend on γ, $B_{zx} = \sqrt{A_{zx}^2 + A_{zy}^2}$ is γ-independent.

[6] The entire powder spectrum can be considered here because the radical has narrow lines, i.e., either DQ or ZQ condition (not both) is calculated here.

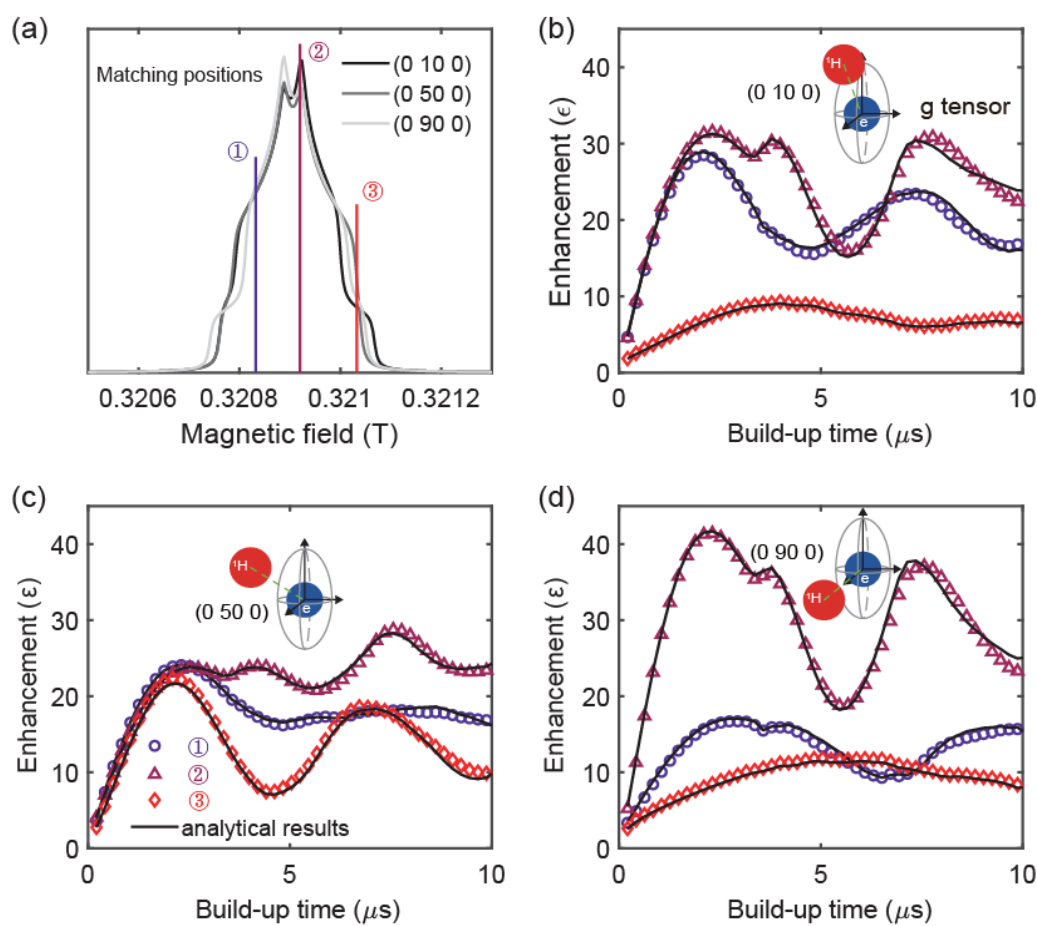

Fig. 4 (a) The calculated electron EPR lineshape of e-$^1$H system with *g*-tensor ($g_x$= 2.0046, $g_y$= 2.0038, $g_z$= 2.0030) and μw frequency 9 GHz. The build-up curves in (b-c) correspond to different μw central frequencies and matching positions labelled in (a). The relative Euler angles between the e-$^1$H dipolar couplings and the *g*-tensor are (0, 10°, 0), (0, 50°, 0), (0°, 90°, 0) for (b-d). The e-$^1$H distance is 4.3 Å.

3.3 The effect of *J*, *d*, and $\Delta B_{zx}$ on the CE-DNP enhancement

It is known in the literature that the *e-e* interactions—exchange interaction *J* and dipolar coupling *d*—play crucial roles in affecting CE DNP performance.[50–52] In particular, *Equbal* et al. noted from numerically simulated results that the CE radicals should have $J/d > 1.25$ for an efficient MAS DNP transfer.[39,53] We show that the phenomenological finding can be explained by inspecting the orientation-dependent build-up rate $\omega_{CE}(\alpha,\beta,\gamma) \propto (2J + d(\alpha,\beta,\gamma))$ (Eq. 29),[7] which requires |2J +

---
[7] $\alpha, \beta, \gamma$ are the relative Euler angles between the crystal to the lab frame.

$d(\alpha,\beta,\gamma)| > 0$ so that all crystallites have non-zero build-up rates even if the CE matching conditions are fulfilled, i.e.:

$$|2J + d(\alpha,\beta,\gamma)| > 0$$

$$\text{Either } 2J - d_{ee} > 0 \rightarrow J/d_{ee} > 1/2$$

$$\text{or } 2J + d_{ee}/2 < 0 \rightarrow J/d_{ee} < -1/4 \tag{35}$$

$$\text{where } d = \frac{1}{2}d_{ee}(1 - 3\cos^2\beta) \text{ and } d_{ee} = \mu_0 \gamma_e^2 \hbar / 4\pi r_{ee}^3$$

Hence, enforcing $|J/d_{ee}| > 1/2$ would ensure that no crystallite will have an instantaneous $\omega_{CE} = 0$ for any orientation in a rotor period. Moreover, having $|J/d_{ee}| \gg 1/2$ would guarantee that the buildup rate $\omega_{CE}$ is moderately higher than a certain threshold, thereby yielding a faster and more efficient DNP transfer. Nevertheless, the $|J/d_{ee}|$ ratio cannot be increased indefinitely, or else it might have a deleterious effect. For instance, the simplified CE condition $\omega_{0I} \sim \pm \Delta\omega$ is no longer applicable for the strong $J$ case, and the full CE matching condition $\omega_{0I} = \pm\sqrt{(2J+d)^2 + \Delta\omega^2}$ (Eq. 28) dictates that the CE condition can never be fulfilled if $(2J + d) > \omega_{0I}$. We will not discuss this further here as the actual CE MAS DNP scenario will be more complex when *g*-anisotropy is considered, and it is beyond the scope of this work. Nevertheless, we emphasize that the derived matching conditions and buildup rates remain valid for static and MAS cases.

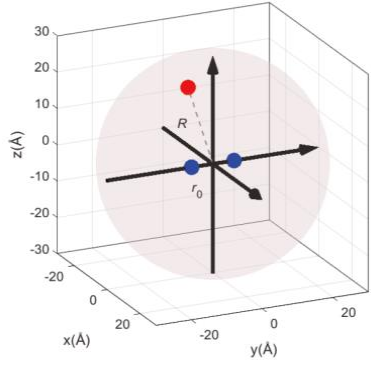

Fig. 5 A three-spin $e_1$-$e_2$-$^1$H model with the coordinates of the spins given by $\tilde{r}_{e1} = (0, r_0, 0)$, $\tilde{r}_{e2} = (0, -r_0, 0)$, $\tilde{r}_{1H} = R(\sin\theta\cos\phi, \sin\theta\sin\phi, \cos\theta)$; The distances are $r_0 = 6$ Å and $R = 30$ Å. The two electrons (blue sphere) have fixed positions, while the angles $\theta$ and $\phi$ of the $^1$H atom (red sphere) are varied. Other spin parameters include $g_{e1}$=2.0000, $g_{e2}$=2.0030 (isotropic) satisfying the CE matching condition, $J = 0$ Hz, $T_{1e}$=1 ms, $T_{2e}$=5 μs, $B_0$=5 T, and microwaves of $\omega_{1S}/2\pi = 4$ MHz is applied on $e_1$.

It is known in the literature that the nature of the $^1$H nuclei close to the electron plays a significant role in DNP. For instance, there exists a sweet spot in which the e-$^1$H distance should be short for efficient DNP contact/transfer but larger than the spin diffusion barrier so that the polarization can be distributed across the bulk sample. Recent literature has reported that the size of this sweet spot is ~3-6 Å away from the radical.[54–56] Although our unified theory and a simple three-spin model here will not be sufficient to treat the spin diffusion barrier issue, we plan to analyze the role of the differential hyperfine interaction, $\Delta B_{zx}$, in mediating CE DNP.

We set up an $e_1$-$e_2$-$^1$H three-spin system in which the two electrons are separated by $2r_0$ = 12 Å, and the $^1$H nucleus is on a spherical shell with a radius $R = 30$ Å away from the origin (Fig. 5). For this study, the two electrons are fixed in position, but the angles $(\theta, \phi)$ will be varied. Figure 6 shows the simulated ε (Fig. 6a) and calculated $\Delta B_{zx}$ (Fig. 6b) for various $^1$H's locations on the $R = 30$ Å shell (or different $\theta$ and $\phi$ angles). The two profiles are very similar and imply a correlation between ε and $\Delta B_{zx}$. To corroborate the results, the data from these two plots are sampled and replotted in Fig. 6c, showing the relation of ε against $\Delta B_{zx}$, which shows clearly that high $\Delta B_{zx}$ values yield high ε, and the converse is also true. These findings confirm the $\omega_{CE} \propto \Delta B_{zx}$ relation (Eq. 29) derived from our unified theory.

Moreover, it is intriguing that there are some blind spots with minimum ε in the *z*=0 plane (equator) and some local spots (see red arrows in Fig. 6b). To further understand this phenomenon, we will first write down the expressions of $\Delta B_{zx}$ and $A^{(i)}_{zx,y}$ in this spin system:

$$\Delta B_{zx} = \sqrt{\left(A^{(1)}_{zx} - A^{(2)}_{zx}\right)^2 + \left(A^{(1)}_{zy} - A^{(2)}_{zy}\right)^2}$$

$$A^{(i)}_{zx} = -\frac{3}{2} d_i \sin 2\theta_i \cos \phi_i \tag{36}$$

$$A^{(i)}_{zy} = -\frac{3}{2} d_i \sin 2\theta_i \sin \phi_i,$$

where and $d_i = \mu_0 \gamma_e \gamma_I \hbar / 4\pi r^3_{e_i H}$ is the electron-nucleus dipolar coupling, $\theta_i$ is the angle between the dipole and the external $B_0$ field, and $\phi_i$ is the azimuth angle. Two solutions are obtained by setting $\Delta B_{zx} = 0$ (Eq. 35): (1) $A^{(1)}_{zx} = A^{(2)}_{zx} = A^{(1)}_{zy} = A^{(2)}_{zy} = 0$ or (2) $A^{(1)}_{zx} = A^{(2)}_{zx}$ and $A^{(1)}_{zy} = A^{(2)}_{zy}$. Indeed, the solution of the first case is $\theta = \pi/2$ (or z = 0 equator). By solving the second case:

$$\Delta B_{zx} = 0 \begin{cases} \dfrac{R \sin\theta + r_0}{(R^2 + 2Rr_0 \sin\theta + r_0^2)^{5/2}} = \dfrac{R \sin\theta - r_0}{(R^2 - 2Rr_0 \sin\theta + r_0^2)^{5/2}} \\ \\ \cos\phi = 0 \end{cases}, \tag{37}$$

one obtains the positions of the ¹H atoms are (0, ±14.06, ±26.50) Å or $\theta$ =27.9° (Fig. 6b), which are as expected. This is interesting because, in the SE DNP case, the ε = 0 blind spot would be at the magic angle $\theta = 54.7°$, where the dipolar coupling is also zero. However, our unified theory has successfully revealed that this is not the case in the CE, and the blind spots are at the regions where the two hyperfine fields exactly equal (or differential hyperfine interaction $\Delta B_{zx} = 0$)—a phenomenon that has not yet been discussed in the literature.

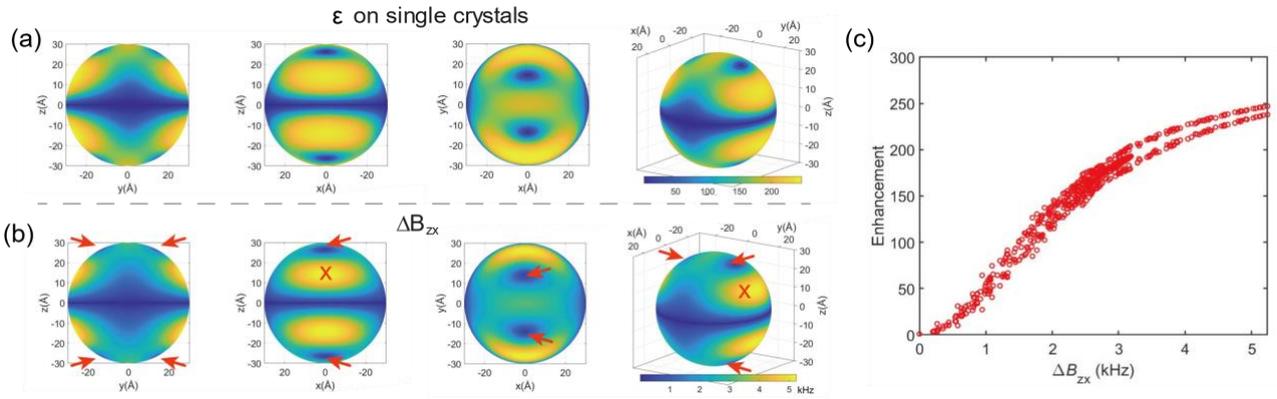

Fig. 6 (a) Simulated CE-DNP enhancement $\varepsilon$ and (b) calculated $\Delta B_{zx}$ for single crystals on a three-spin system shown in Fig. 5. (b) The $\Delta B_{zx}=0$ and the maximum $\Delta B_{zx}$ regions are labelled by red arrows red crosses, respectively. (c) Plot of $\varepsilon$ against $\Delta B_{zx}$ using resampled data from (a) and (b).

For powdered samples, the general features exhibited by the simulated $\varepsilon$ (Fig. 7a) and powder-averaged $\langle \Delta B_{zx} \rangle$ (Fig. 7b) are similar, i.e., the high $\varepsilon$ regions are well reflected by the high calculated $\langle \Delta B_{zx} \rangle$ values. However, some differences are also noted: (1) there are starker contrasts between the annular rings (red crosses in Fig. 7a) and (2) the asymmetry between the two $z$-hemispheres is not observed in the $\langle \Delta B_{zx} \rangle$ plot (Fig. 7b). To address issue (1), we incorporated the effects of CE mismatches due to the orientation-dependent dipolar couplings (see Section 3.1) and calculated $\langle \omega'_{CE} \cos^2 \zeta_{CE} \rangle$. The resulting $\langle \omega'_{CE} \cos^2 \zeta_{CE} \rangle$ plot (Fig. 7c) for the powder subset shows a much better agreement with simulated $\varepsilon$. The strength of the unified theoretical framework allowing direct adaptation of the SE scenario for CE is again exemplified here. For issue (2), the $\varepsilon$ asymmetry between the +y and -y hemispheres (Fig. 6a, Fig. 7a) can be explained by the size of the exchange interaction $J$ (2), selective excitation on one of the two electrons, and relaxation effects. As the issue is multifaceted and complex, we will not discuss it further.

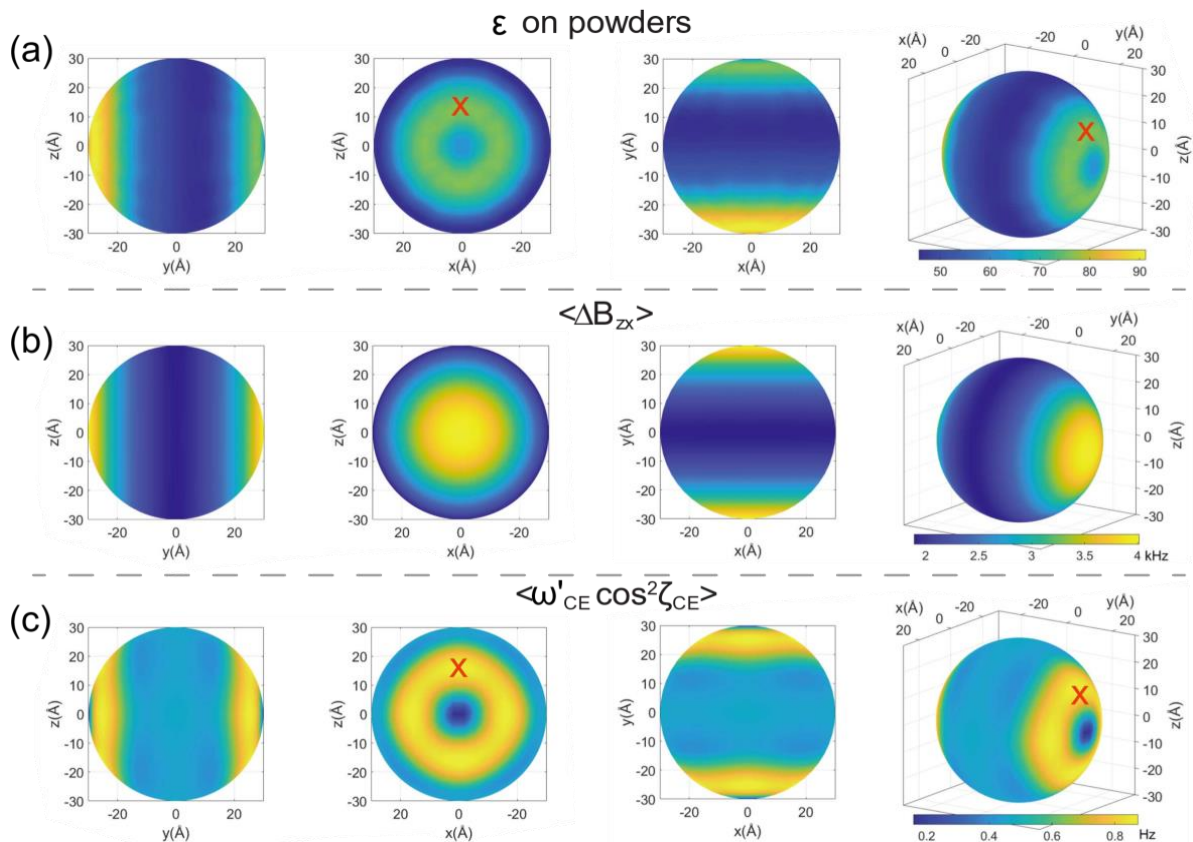

Fig. 7 (a) Simulated CE-DNP enhancement ε, (b) calculated $\Delta B_{zx}$ and (c) $\omega'_{CE} \cos^2 \zeta_{CE}$ for powders on a three-spin system shown in Fig. 5c. The maximum enhancement and $\omega'_{CE} \cos^2 \zeta_{CE}$ regions are labelled by red crosses.

We have demonstrated here that our unified theory has shed new light on the role of differential hyperfine interaction $\Delta B_{zx}$ (or the position of nearest $^1$H) in dictating the CE-DNP performance. These findings could be exploited to design more efficient biradicals by avoiding these zero-enhancement blind spots—either by optimizing the linkers or deuterating the $^1$H's at those regions.

4  Conclusion

We have provided an analytical description for CP, NOVEL, SE, and CE mechanisms using the same unified theoretical framework. Not only the use of fictitious spin-1/2 operators combined with average Hamiltonian theory provides an easy-to-understand and intuitive explanation for the polarization-transfer mechanisms, but it also sheds new light on fundamental DNP mechanisms. For instance, we show that the DNP build-up curves should also feature transient oscillations, which could be exploited to extract crucial structural information in metal-doped paramagnetic biomolecules or materials, i.e., we would like to extend the DNP applications beyond just hyperpolarization. Moreover, the realization that SE and NOVEL have $\gamma$-independent DNP performances, and that an inverted electron polarization could generate higher DNP enhancement than a simple saturation scheme in CE could motivate further development of new (pulsed) DNP sequences in the future. Besides that, our theory sheds light on the roles of exchange interaction ($J$) and $\Delta B_{zx}$ in CE. In particular, the CE matching conditions helped explain the phenomenological finding of a good *J/D* ratio for efficient CE in the literature. Moreover, the theory also highlighted the importance of the differential hyperfine interaction $\Delta B_{zx}$, which is directly correlated to the CE enhancement factors. These results can potentially be exploited for designing more efficient biradicals. Additionally, although our study here is performed only on the static case, our unified theory remains valid and can be extended for the MAS case if needed. At last, we hope that our presented findings here could stimulate an experimental effort in verifying our theory and numerical results—when high-power pulsed microwave devices at high fields become available in the future.

# Appendix

## 1 Matching conditions

The simplified form of the Hamiltonian in the interaction frame for NOVEL (Eq. 11) and SE (Eq. 19) are:

NOVEL:

$$\widehat{\widetilde{\mathcal{H}}}(t) = -\frac{B_{zx}}{2}\left[\left(\hat{S}_x\hat{I}_x - \hat{S}_y\hat{I}_y\right)\cos(\omega_{1S} - \omega_{0I})t - \left(\hat{S}_x\hat{I}_y + \hat{S}_y\hat{I}_x\right)\sin(\omega_{1S} - \omega_{0I})t\right]$$
$$-\frac{B_{zx}}{2}\left[\left(\hat{S}_x\hat{I}_x + \hat{S}_y\hat{I}_y\right)\cos(\omega_{1S} + \omega_{0I})t + \left(\hat{S}_x\hat{I}_y - \hat{S}_y\hat{I}_x\right)\sin(\omega_{1S} + \omega_{0I})t\right]$$
$$-A_{zz}\left(\cos\omega_{1S}\,\hat{S}_x\hat{I}_z - \sin\omega_{1S}t\,\hat{S}_y\hat{I}_z\right).$$

SE: (S1)

$$\widehat{\widetilde{\mathcal{H}}}(t) = A_{zz}\cos\theta\,\hat{S}_z\hat{I}_z$$
$$-\frac{B_{zx}\sin\theta}{2}\left[\left(\hat{S}_x\hat{I}_x - \hat{S}_y\hat{I}_y\right)\cos(\omega_{eff} - \omega_{0I})t - \left(\hat{S}_x\hat{I}_y + \hat{S}_y\hat{I}_x\right)\sin(\omega_{eff} - \omega_{0I})t\right]$$
$$-\frac{B_{zx}\sin\theta}{2}\left[\left(\hat{S}_x\hat{I}_x + \hat{S}_y\hat{I}_y\right)\cos(\omega_{eff} + \omega_{0I})t + \left(\hat{S}_x\hat{I}_y - \hat{S}_y\hat{I}_x\right)\sin(\omega_{eff} + \omega_{0I})t\right]$$
$$-A_{zz}\sin\theta\left(\hat{S}_x\hat{I}_z\cos\omega_{eff}t - \hat{S}_y\hat{I}_z\sin\omega_{eff}t\right)$$
$$+B_{zx}\cos\theta\left(\hat{S}_z\hat{I}_x\cos\omega_{0I}t + \hat{S}_z\hat{I}_y\sin\omega_{0I}t\right),$$

It is evident that the matching condition are $\omega_{1S} = \pm\omega_{0I}$ and $\omega_{eff} = \pm\omega_{0I}$ for NOVEL and SE cases, respectively.

## 2 Examples of asymmetry enhancement plot

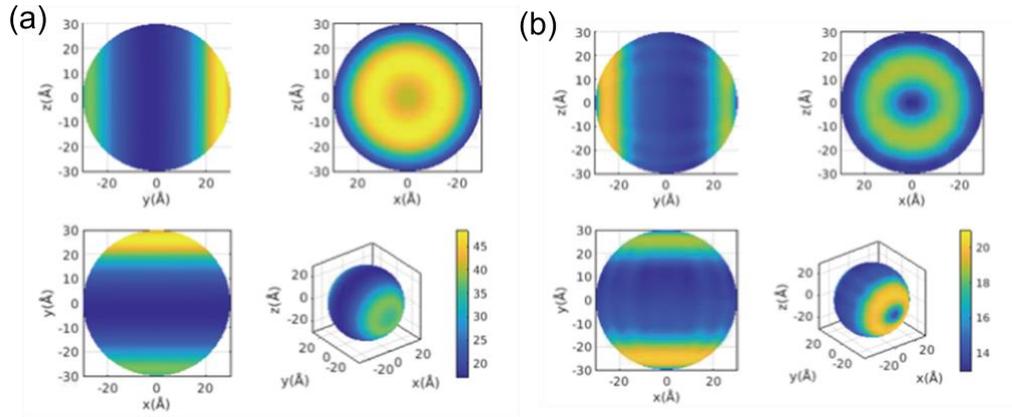

Fig. S1 Simulated CE-DNP enhancement in powders with (a) $J$ = 9 MHz (b) and $-9$ MHz. One can reverse the $\varepsilon$ asymmetry by reversing the sign of exchange interaction.

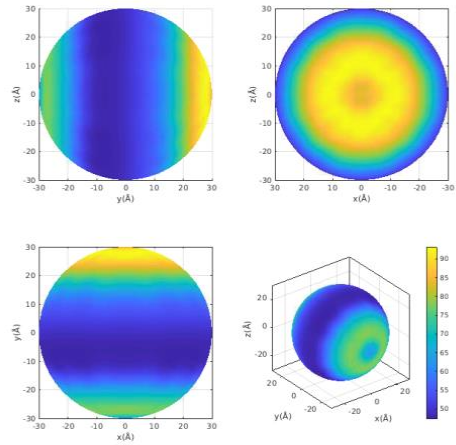

Fig. S2 Simulated CE-DNP enhancement in powders. One can reverse the $\varepsilon$ asymmetry (with respect to Fig. 6d) by saturating the second $e_2$ (0 -6 0) instead of the first electron $e_1$.

## 3 Definition of fictitious spin-1/2 operators

| Original operators | $\frac{1}{2}(\hat{S}_z + \hat{I}_z)$ | $\frac{1}{2}(\hat{S}_z - \hat{I}_z)$ | $\hat{S}_x\hat{I}_y + \hat{S}_x\hat{I}_y$ | $\hat{S}_y\hat{I}_x - \hat{S}_x\hat{I}_y$ | $\hat{S}_x\hat{I}_x - \hat{S}_y\hat{I}_y$ | $\hat{S}_x\hat{I}_x + \hat{S}_y\hat{I}_y$ |
|---|---|---|---|---|---|---|
| Fictitious operators | $\hat{S}_z^\Sigma$ | $\hat{S}_z^\Delta$ | $\hat{S}_y^\Sigma$ | $\hat{S}_y^\Delta$ | $\hat{S}_x^\Sigma$ | $\hat{S}_x^\Delta$ |

# References

[1]     S.R. Hartmann, E.L. Hahn, Nuclear Double Resonance in the Rotating Frame, Phys. Rev. 128 (1962) 2042–2053. https://doi.org/10.1103/PhysRev.128.2042.

[2]     A. Abragam, M. Goldman, Principles of dynamic nuclear polarisation, Reports Prog. Phys. 41 (1978) 395–467. https://doi.org/10.1088/0034-4885/41/3/002.

[3]     A.S. Lilly Thankamony, J.J. Wittmann, M. Kaushik, B. Corzilius, Dynamic nuclear polarization for sensitivity enhancement in modern solid-state NMR, Prog. Nucl. Magn. Reson. Spectrosc. 102–103 (2017) 120–195. https://doi.org/10.1016/j.pnmrs.2017.06.002.

[4]     L. Müller, A. Kumar, T. Baumann, R.R. Ernst, Transient Oscillations in NMR Cross-Polarization Experiments in Solids, Phys. Rev. Lett. 32 (1974) 1402–1406. https://doi.org/10.1103/PhysRevLett.32.1402.

[5]     O.W. Sørensen, G.W. Eich, M.H. Levitt, G. Bodenhausen, R.R. Ernst, Product operator formalism for the description of NMR pulse experiments, Prog. Nucl. Magn. Reson. Spectrosc. 16 (1984) 163–192. https://doi.org/10.1016/0079-6565(84)80005-9.

[6]     V. Ladizhansky, S. Vega, Polarization transfer dynamics in Lee–Goldburg cross polarization nuclear magnetic resonance experiments on rotating solids, J. Chem. Phys. 112 (2000) 7158. https://doi.org/10.1063/1.481281.

[7]     C.A. Fyfe, A.R. Lewis, J.-M. Chézeau, A comparison of NMR distance determinations in the solid state by cross polarization, REDOR, and TEDOR techniques, (1999).

[8]     J. Giraudet, M. Dubois, A. Hamwi, W.E.E. Stone, P. Pirotte, F. Masin, Solid-State NMR (19F and 13C) Study of Graphite Monofluoride (CF)n: 19F Spin−Lattice Magnetic Relaxation and 19F/13C Distance Determination by Hartmann−Hahn Cross Polarization, J. Phys. Chem. B. 109 (2004) 175–181. https://doi.org/10.1021/JP046833J.

[9]     A. Abragam, W.G. Proctor, Une Nouvelle Methode De Polarisation Dynamique Des Noyaux Atomiques Dans Les Solides, Comptes Rendus Hebdo- Madaires Des Seances L Acad. Des Sci. 246 (1958) 2253–2256.

[10]    C.F. Hwang, D.A. Hill, New effect in dynamic polarization, Phys. Rev. Lett. 18 (1967) 110–112. https://doi.org/10.1103/PhysRevLett.18.110.

[11]    C.F. Hwang, D.A. Hill, Phenomenological Model for the New Effect in Dynamic Polarization, Phys. Rev. Lett. 19 (1967) 1011–1014. https://doi.org/10.1103/PhysRevLett.19.1011.

[12]    K.R. Thurber, R. Tycko, Theory for cross effect dynamic nuclear polarization under magic-angle spinning in solid state nuclear magnetic resonance: The importance of level crossings, J. Chem. Phys. 137 (2012) 084508. https://doi.org/10.1063/1.4747449.

[13]    W.T. Wenckebach, The Solid Effect, Appl. Magn. Reson. 34 (2008) 227–235. https://doi.org/10.1007/s00723-008-0121-9.

[14]    K.-N. Hu, G.T. Debelouchina, A.A. Smith, R.G. Griffin, Quantum mechanical theory of dynamic nuclear

polarization in solid dielectrics., J. Chem. Phys. 134 (2011) 125105. https://doi.org/10.1063/1.3564920.

[15] J. van Houten, W.T. Wenckebach, N.J. Poulis, A study of the thermal contact between the nuclear Zeeman system and the electron dipole-dipole interaction system, Phys. B+C. 92 (1977) 210–220. https://doi.org/10.1016/0378-4363(77)90021-3.

[16] M. Rosay, M. Blank, F. Engelke, Instrumentation for solid-state dynamic nuclear polarization with magic angle spinning NMR., J. Magn. Reson. 264 (2016) 88–98. https://doi.org/10.1016/j.jmr.2015.12.026.

[17] J.H. Ardenkjaer-Larsen, B. Fridlund, A. Gram, G. Hansson, L. Hansson, M.H. Lerche, R. Servin, M. Thaning, K. Golman, Increase in signal-to-noise ratio of > 10,000 times in liquid-state NMR, Proc. Natl. Acad. Sci. 100 (2003) 10158–10163. https://doi.org/10.1073/pnas.1733835100.

[18] L.R. Becerra, G.J. Gerfen, R.J. Temkin, D.J. Singel, R.G. Griffin, Dynamic nuclear polarization with a cyclotron resonance maser at 5 T, Phys. Rev. Lett. 71 (1993) 3561–3564. https://doi.org/10.1103/PhysRevLett.71.3561.

[19] K.O. Tan, S. Jawla, R.J. Temkin, R.G. Griffin, Pulsed Dynamic Nuclear Polarization, in: EMagRes, 2019: pp. 339–352. https://doi.org/10.1002/9780470034590.emrstm1551.

[20] A. Henstra, P. Dirksen, J. Schmidt, W.T. Wenckebach, Nuclear spin orientation via electron spin locking (NOVEL), J. Magn. Reson. 77 (1988) 389–393. https://doi.org/10.1016/0022-2364(88)90190-4.

[21] H. Brunner, R.H. Fritsch, K.H. Hausser, Notizen: Cross Polarization in Electron Nuclear Double Resonance by Satisfying the Hartmann-Hahn Condition, Zeitschrift Für Naturforsch. A. 42 (1987) 1456–1457. https://doi.org/10.1515/zna-1987-1217.

[22] M.M. Maricq, J.S. Waugh, NMR in rotating solids, J. Chem. Phys. 70 (1979) 3300. https://doi.org/10.1063/1.437915.

[23] U. Haeberlen, J. Waugh, Coherent Averaging Effects in Magnetic Resonance, Phys. Rev. 175 (1968) 453–467. https://doi.org/10.1103/PhysRev.175.453.

[24] A. Henstra, W.T. Wenckebach, The theory of nuclear orientation via electron spin locking (NOVEL), Mol. Phys. 106 (2008) 859–871. https://doi.org/10.1080/00268970801998262.

[25] T. V. Can, J.J. Walish, T.M. Swager, R.G. Griffin, Time domain DNP with the NOVEL sequence, J. Chem. Phys. 143 (2015) 054201. https://doi.org/10.1063/1.4927087.

[26] D.J. van den Heuvel, A. Henstra, T.-S. Lin, J. Schmidt, W.T. Wenckebach, Transient oscillations in pulsed dynamic nuclear polarization, Chem. Phys. Lett. 188 (1992) 194–200. https://doi.org/10.1016/0009-2614(92)90008-B.

[27] S.K. Jain, G. Mathies, R.G. Griffin, Off-resonance NOVEL, J. Chem. Phys. 147 (2017) 164201. https://doi.org/10.1063/1.5000528.

[28] G. Mathies, S. Jain, M. Reese, R.G. Griffin, Pulsed Dynamic Nuclear Polarization with Trityl Radicals., J. Phys. Chem. Lett. 7 (2016) 111–116. https://doi.org/10.1021/acs.jpclett.5b02720.

[29] Y. Hovav, A. Feintuch, S. Vega, Theoretical aspects of dynamic nuclear polarization in the solid state - The solid effect, J. Magn. Reson. 207 (2010) 176–189. https://doi.org/10.1016/j.jmr.2010.10.016.

[30] B. Corzilius, A.A. Smith, R.G. Griffin, Solid effect in magic angle spinning dynamic nuclear polarization., J. Chem. Phys. 137 (2012) 54201. https://doi.org/10.1063/1.4738761.

[31] K.-N. Hu, V.S. Bajaj, M. Rosay, R.G. Griffin, High-frequency dynamic nuclear polarization using mixtures of TEMPO and trityl radicals, J. Chem. Phys. 126 (2007) 44512. https://doi.org/10.1063/1.2429658.

[32] C. Song, K.-N. Hu, C.-G. Joo, T.M. Swager, R.G. Griffin, TOTAPOL: A Biradical Polarizing Agent for Dynamic Nuclear Polarization Experiments in Aqueous Media, J. Am. Chem. Soc. 128 (2006) 11385–11390. https://doi.org/10.1021/ja061284b.

[33] K.-N. Hu, H. Yu, T.M. Swager, R.G. Griffin, Dynamic nuclear polarization with biradicals., J. Am. Chem. Soc. 126 (2004) 10844–10845. https://doi.org/10.1021/ja039749a.

[34] D. Mance, P. Gast, M. Huber, M. Baldus, K.L. Ivanov, The magnetic field dependence of cross-effect dynamic nuclear polarization under magic angle spinning, J. Chem. Phys. 142 (2015) 234201. https://doi.org/10.1063/1.4922219.

[35] A. Equbal, S. Jain, Y. Li, K. Tagami, X. Wang, S. Han, Role of electron spin dynamics and coupling network in designing dynamic nuclear polarization, Prog. Nucl. Magn. Reson. Spectrosc. (2021) شماره 8; 117--99 ص. https://doi.org/10.1016/j.pnmrs.2021.05.003.

[36] Y. Hovav, O. Levinkron, A. Feintuch, S. Vega, Theoretical Aspects of Dynamic Nuclear Polarization in the Solid State: The Influence of High Radical Concentrations on the Solid Effect and Cross Effect Mechanisms, Appl. Magn. Reson. 43 (2012) 21–41. https://doi.org/10.1007/s00723-012-0359-0.

[37] F. Mentink-Vigier, Ü. Akbey, H. Oschkinat, S. Vega, A. Feintuch, Theoretical aspects of Magic Angle Spinning - Dynamic Nuclear Polarization, J. Magn. Reson. 258 (2015) 102–120. https://doi.org/10.1016/j.jmr.2015.07.001.

[38] Y. Hovav, A. Feintuch, S. Vega, Theoretical aspects of dynamic nuclear polarization in the solid state - The cross effect, J. Magn. Reson. 214 (2012) 29–41. https://doi.org/10.1016/j.jmr.2011.09.047.

[39] A. Equbal, A. Leavesley, S.K. Jain, S. Han, Cross-Effect Dynamic Nuclear Polarization Explained: Polarization, Depolarization, and Oversaturation, J. Phys. Chem. Lett. (2019). https://doi.org/10.1021/ACS.JPCLETT.8B02834.

[40] V.K. Michaelis, A.A. Smith, rn Corzilius, O. Haze, T.M. Swager, R.G. Griffin, High-Field 13 C Dynamic Nuclear Polarization with a Radical Mixture, J. Am. Chem. Soc. 12 (2013) 5. https://doi.org/10.1021/ja312265x.

[41] G. Mathies, M.A. Caporini, V.K. Michaelis, Y. Liu, K.N. Hu, D. Mance, J.L. Zweier, M. Rosay, M. Baldus, R.G. Griffin, Efficient Dynamic Nuclear Polarization at 800 MHz/527 GHz with Trityl-Nitroxide Biradicals, Angew. Chemie - Int. Ed. 54 (2015) 11770–11774. https://doi.org/10.1002/anie.201504292.

[42] F. Mentink-Vigier, V. Rane, T. Dubroca, K. Kundu, Spinning Driven Dynamic Nuclear Polarization with Optical Pumping, (2022). https://arxiv.org/abs/2201.05703v2 (accessed January 27, 2022).

[43] G. Jeschke, A new mechanism for chemically induced dynamic nuclear polarization in the solid state, J. Am. Chem. Soc. 120 (1998) 4425–4429. https://doi.org/10.1021/ja973744u.

[44]   Spinach, (n.d.). http://spindynamics.org/group/?page_id=12.

[45]   R.I. Hunter, P.A.S. Cruickshank, D.R. Bolton, P.C. Riedi, G.M. Smith, High power pulsed dynamic nuclear polarisation at 94 GHz, Phys. Chem. Chem. Phys. 12 (2010) 5752. https://doi.org/10.1039/c002251a.

[46]   G. Pileio, M. Concistrè, N. McLean, A. Gansmüller, R.C.D. Brown, M.H. Levitt, Analytical theory of gamma-encoded double-quantum recoupling sequences in solid-state nuclear magnetic resonance., J. Magn. Reson. 186 (2007) 65–74. https://doi.org/10.1016/j.jmr.2007.01.009.

[47]   K.O. Tan, A.B. Nielsen, B.H. Meier, M. Ernst, Broad-Band DREAM Recoupling Sequence, J. Phys. Chem. Lett. 5 (2014) 3366–3372. https://doi.org/10.1021/jz501703e.

[48]   S. Jain, M. Bjerring, N.C. Nielsen, Efficient and Robust Heteronuclear Cross-Polarization for High-Speed-Spinning Biological Solid-State NMR Spectroscopy, J. Phys. Chem. Lett. 3 (2012) 703–708. https://doi.org/10.1021/jz3000905.

[49]   N.C. Nielsen, H. Bildsøe, H.J. Jakobsen, M.H. Levitt, Double-quantum homonuclear rotary resonance: Efficient dipolar recovery in magic-angle spinning nuclear magnetic resonance, J. Chem. Phys. 101 (1994) 1805. https://doi.org/10.1063/1.467759.

[50]   K. Tagami, A. Equbal, I. Kaminker, B. Kirtman, S. Han, Biradical rotamer states tune electron J coupling and MAS dynamic nuclear polarization enhancement, Solid State Nucl. Magn. Reson. 101 (2019) 12–20. https://doi.org/10.1016/J.SSNMR.2019.04.002.

[51]   F. Mentink-Vigier, S. Vega, G. De Paëpe, Fast and accurate MAS–DNP simulations of large spin ensembles, Phys. Chem. Chem. Phys. 19 (2017) 3506–3522. https://doi.org/10.1039/C6CP07881H.

[52]   F. Mentink-Vigier, I. Marin-Montesinos, A.P. Jagtap, T. Halbritter, J. Van Tol, S. Hediger, D. Lee, S.T. Sigurdsson, G. De Paëpe, Computationally Assisted Design of Polarizing Agents for Dynamic Nuclear Polarization Enhanced NMR: The AsymPol Family, J. Am. Chem. Soc. 140 (2018) 11013–11019. https://doi.org/10.1021/JACS.8B04911/SUPPL_FILE/JA8B04911_SI_001.PDF.

[53]   A. Equbal, K. Tagami, S. Han, Balancing Dipolar and Exchange Coupling in Biradicals to Maximize Cross Effect Dynamic Nuclear Polarization, Phys. Chem. Chem. Phys. 19 (2020). https://doi.org/10.1039/D0CP02051F.

[54]   K.O. Tan, M. Mardini, C. Yang, J.H. Ardenkjær-Larsen, R.G. Griffin, Three-spin solid effect and the spin diffusion barrier in amorphous solids, Sci. Adv. 5 (2019) eaax2743. https://doi.org/10.1126/sciadv.aax2743.

[55]   Q. Stern, S.F. Cousin, F. Mentink-Vigier, A.C. Pinon, S.J. Elliott, O. Cala, S. Jannin, Direct observation of hyperpolarization breaking through the spin diffusion barrier, Sci. Adv. 7 (2021) 1–14. https://doi.org/10.1126/sciadv.abf5735.

[56]   S.K. Jain, C.-J. Yu, C.B. Wilson, T. Tabassum, D.E. Freedman, S. Han, Dynamic Nuclear Polarization with Vanadium(IV) Metal Centers, Chem. (2020) 86–1226. https://doi.org/10.1016/j.chempr.2020.10.021.